\documentclass[review]{elsarticle}

\usepackage{hyperref,amsmath,amssymb,bm,amsfonts,times}
\usepackage{graphicx}
\usepackage{textcomp}
\usepackage{xcolor}
\usepackage{subfigure}
\usepackage{epsfig}
\usepackage{multirow}
\usepackage{algorithm}
\usepackage{algorithmicx}
\usepackage{algpseudocode}
\usepackage{epstopdf}
\usepackage{array}
\usepackage{stfloats}
\usepackage{setspace}
\usepackage{enumerate}
\usepackage{geometry}
\usepackage[toc,page]{appendix}
\journal{Digital Signal Processing}

\begin{document}

\begin{frontmatter}


\title{ Proportionate Recursive Maximum Correntropy Criterion Adaptive Filtering Algorithms and their Performance Analysis}
\author[a,b]{Zhen Qin}
\ead{qin.660@osu.edu}
\author[c]{Jun Tao\corref{cor1}}
\cortext[cor1]{Corresponding authors.}
\ead{jtao@seu.edu.cn}
\author[d]{Le Yang}
\ead{le.yang@cantebury.ac.nz}
\author[e]{Ming Jiang}
\ead{jiangm@pcl.ac.cn}
\address[a]{Department of Computer Science and Engineering, Ohio State University, Columbus, 43210, USA.}
\address[b]{School of Information Science and Engineering, Southeast University, Nanjing, 210096, China.}
\address[c]{Key Laboratory of Underwater Acoustic Signal Processing of the Ministry of Education, \\School of Information Science and Engineering, Southeast University, Nanjing 210096, China.}
\address[d]{Department of ECE, University of Canterbury, Christchurch, 8020, New Zealand.}
\address[e]{Pengcheng Lab, Shenzhen, 518066, China.}
\begin{abstract}

The maximum correntropy criterion (MCC) has been employed to design outlier-robust adaptive filtering algorithms, among which the recursive MCC (RMCC) algorithm is a typical one. Motivated by the success of our recently proposed proportionate recursive least squares (PRLS) algorithm for sparse system identification, we propose to introduce the proportionate updating (PU) mechanism into the RMCC, leading to two sparsity-aware RMCC algorithms: the proportionate recursive MCC (PRMCC) algorithm and the combinational PRMCC (CPRMCC) algorithm. The CPRMCC is implemented as an adaptive convex combination of two PRMCC filters.
For PRMCC, its stability condition and mean-square performance were analyzed. Based on the analysis, optimal parameter selection in nonstationary environments was obtained. Performance study of CPRMCC was also provided and showed that the CPRMCC performs at least as well as the better component PRMCC filter in steady state. 
Numerical simulations of sparse system identification corroborate the advantage of proposed algorithms as well as the validity of theoretical analysis.
\end{abstract}
\end{frontmatter}

\vspace{-0.2cm}
\section{Introduction}
The minimum mean square error (MMSE) and least squares (LS) criteria, have been widely adopted in the design of adaptive filtering algorithms, leading to e.g. the least mean squares (LMS) and recursive least squares (RLS) filters. These classic algorithms have attained great success in many applications such as channel equalization \cite{Duan18}, noise cancellation \cite{Yang11}, system modeling \cite{Pelekanakis14}, just to name a few. However, the MMSE/LS criteria are known to be sensitive to outliers \cite{Chen17}. Outliers are common in various contexts, such as sensor calibration \cite{LiTSP17} (due to sensor failure), face recognition \cite{Torre03} (owing to self-shadowing, specularity, or brightness saturation), and others. In contrast, the emerging maximum correntropy criterion (MCC) is more robust to outliers.  

Many MCC adaptive filtering algorithms have been developed \cite{Singh09,Chen14,Chen16,Singh10,Chen15,Qian18,Ma13}. They can be classified into two categories: stochastic gradient type algorithms \cite{Singh09,Chen14,Chen16} and recursive MCC (RMCC) type algorithms \cite{Singh10,Chen15,Qian18,Ma13}. The design of RMCC is somewhat similar to that of the RLS. In \cite{Singh09}, an MCC algorithm was proposed for the environment with impulsive noise and its theoretical excess mean square error (MSE) was derived using the energy conservation principle \cite{Chen14}. In \cite{Chen16}, a generalized Gaussian density (GGD) cost function was employed to replace the Gaussian kernel used in original MCC algorithm, leading to a generalized maximum correntropy criterion (GMCC) algorithm. Compared with stochastic gradient type algorithms, the RMCC type algorithms can achieve superlinear convergence \cite{Singh10,Chen15,Qian18}.

Motivated by the success of sparsity-aware MMSE/LS adaptive filtering algorithms \cite{Chen09,Gu09,Babadi10,Eksioglu11,Eksioglu111}, the design of sparse MCC adaptive filtering algorithms have recently attracted great attentions \cite{Ma15,Ma17,Zhang16,Ma181,Li20}. In \cite{Ma15}, a correntropy induced metric (CIM) MCC (CIMMCC) algorithm has been designed by exploiting the CIM to approximate the $l_0$ norm in the cost function. In \cite{Ma17,Zhang16}, the $l_1$ norm and a general regularization function have been respectively introduced to the RMCC algorithm, leading to faster convergence under sparse systems. In \cite{Ma181,Li20}, the proportionate MCC (PMCC) and hybrid-norm constrained PMCC algorithms were proposed for wide-sense sparse systems \cite{Tao20} and block sparse systems \cite{Liu16}, respectively.

In our recent work \cite{Qin20}, the proportionate updating (PU) mechanism was introduced to the standard RLS, leading to the proportionate recursive least squares (PRLS) algorithm with improved performance under sparse systems. This naturally motivates us to incorporate a similar proportionate matrix to the existing RMCC to establish a proportionate recursive maximum correntropy criterion (PRMCC) algorithm. According to the analysis in \cite{Qin201}, the parameter $\theta$ controlling the trace of the proportionate matrix in the PRLS trades off between the initial convergence and steady-state performance. Specifically, when $\theta\leq N$ ($N$ is the filter length), the steady state of the PRLS is at least as good as the standard RLS. However, its convergence rate may be slower. 
To achieve fast convergence and low steady-state error simultaneously, two methods have been proposed for LMS-type algorithms: variable step approaches \cite{Kwong92,Aboulnasr97,Huang12} and convex combination \cite{Garcia05,Garcia06,Garcia09,Shi14}. As $\theta$ has similar function as the step size in the LMS, we explore the feasibility of further improving the PRMCC.
In particular, we propose an adaptive convex combination of two PRMCC filters, leading to the combinational PRMCC (CPRMCC) algorithm. Theoretical performance analysis of the PRMCC and CPRMCC are then provided. For the PRMCC, its stability condition is first derived based on the Taylor expansion approach. Then, we investigate its steady-state performance and tracking ability, based on which theoretically optimal parameter values can be determined. For the CPRMCC, we prove its performance is at least not worse than the better one of its two components. Simulation results on sparse channel estimation are presented to demonstrate the effectiveness of proposed algorithms.

The rest of this paper is organized as follows.
In Section~\ref{PPRMCCAlgorithm}, we develop the PRMCC and CPRMCC algorithms.
Section~\ref{Perforanalysis} provides the stability condition and analyzes the mean-square performance for the PRMCC in stationary and nonstationary environments. In addition, the superiority of CPRMCC is verified. In Section~\ref{SIMRESULTS}, simulation results are presented and Section~\ref{conclusion} concludes the paper.

{\bf Notation}: We use bold capital letters, boldface lowercase letters, and italic letters, e.g. $\bf{A}$, $\bf{a}$, and $a$ to, respectively, represent matrix, vector, and scalar quantities. The superscript, $(\cdot)^T$, denotes the transpose, and $\text{E}[\cdot]$ represents the statistical expectation.
For a vector $\bf{a}$ of size $N\times 1$, its $l_n$-norm is defined as $||{\bf a}||_n=(\sum_{m=1}^{N}|a_m|^n)^\frac{1}{n}$.
The $\text{tr}({\bf A})$ returns the trace of a matrix ${\bf A}$.

\vspace{0cm}
\section{Sparsity-aware RMCC Adaptive Filtering Algorithms}
\label{PPRMCCAlgorithm}
\vspace{0cm}

Consider a standard system identification setting in the real domain with the following input-output relationship
\begin{eqnarray}
    \label{PPRMCCAlgorithm1}
    d(n)={\bf w}_N^T{\bf x}_N(n)+v(n),
\end{eqnarray}
where ${\bf w}_N=[w_1,w_2,\cdots,w_{N}]^T$ is the $N\times 1$ system impulse response vector to be estimated, and ${\bf x}_N(n)=[x(n),x(n-1),\cdots,x(n-N+1)]^T$ denotes the input vector at time instant $n$. The $d(n)$ and $v(n)$ represent the system output and the additive noise.

Given two random variables $X$ and $Y$, the correntropy is defined as \cite{Liu07}
\begin{eqnarray}
    \label{PPRMCCAlgorithm3}
    V(X,Y)=\text{E}[\kappa(X,Y)]=\int \kappa(x,y)\, dF_{XY}(x,y),
\end{eqnarray}
where $\kappa(\cdot,\cdot)$ is a shift-invariant Mercer kernel, and $F_{XY}(x,y)$ denotes the joint distribution function of $(X,Y)$.
Unless otherwise specified, the following Gaussian kernel is used in this work: 
\begin{eqnarray}
    \label{PPRMCCAlgorithm4}
    \kappa(x,y)=G_{\sigma}(e)=\frac{1}{\sqrt{2\pi}\sigma}{\exp}(-\frac{e^2}{2\sigma^2}),
\end{eqnarray}
where $e=x-y$ and $\sigma>0$ is the kernel bandwidth. The correntropy measures how similar two random variables are within a neighborhood controlled by the kernel width $\sigma$. Specifically, the smaller the value of $\sigma$, the better it can eliminate the impact caused by outliers. While $\sigma\to \infty$, such a metric will reduce to one global measure, i.e., mean square error (MSE).

The RMCC employs the following cost function 
\begin{eqnarray}
    \label{PPRMCCAlgorithm5}
    J_{\rm{RMCC}}(n)=\sum_{i=1}^{n}\lambda^{n-i}{\exp}(-\frac{e^2(i|n)}{2\sigma^2}),
\end{eqnarray}
where $0<\lambda<1$ is a forgetting factor and
\begin{eqnarray}
    \label{error_07Feb23}
    e(i|n)=d(i)-{\bf w}^T_{N}(n){\bf x}_N(i).
\end{eqnarray}
The weight vector ${\bf w}_N(n)=[w_1(n),w_2(n),\cdots,w_{N}(n)]^T$ denotes an estimate of ${\bf w}_N$ at time $n$. According to \cite{Zhang16}, the weight update of RMCC is governed by 
\begin{eqnarray}
    \label{PPRMCCAlgorithm6}
    {\bf w}_N(n)={\bf w}_N(n\!-\!1)+{\bf k}(n)f(e(n|n\!-\!1)),
\end{eqnarray}
where
\begin{eqnarray}
    \label{PPRMCCAlgorithm61}
    f(e(n|n\!-\!1))={\rm{exp}}(-\frac{e^2(n|n\!-\!1)}{2\sigma^2})e(n|n\!-\!1),
\end{eqnarray}
with the {\em a priori} error being $e(n|n-1)=d(n)-{\bf w}^T_{N}(n-1){\bf x}_N(n)$, and the (Kalman) gain vector ${\bf k}(n)$ is
\begin{eqnarray}
    \label{PPRMCCAlgorithm7}
    {\bf k}(n)=\frac{{\bf P}(n\!-\!1){\bf x}_{N}(n)}{\lambda+{\rm{exp}}(-\frac{e^2(n|n\!-\!1)}{2\sigma^2}){\bf x}^{T}_{N}(n){\bf P}(n\!-\!1){\bf x}_{N}(n)}.
\end{eqnarray}
The matrix ${\bf P}(n-1)$ can be iteratively computed using 
\begin{eqnarray}
    \label{PPRMCCAlgorithm8}
    {\bf P}(n-1)=\lambda^{-1}\left[{\bf P}(n-2)-{\rm{exp}}(-\frac{e^2(n\!-\!1|n\!-\!1)}{2\sigma^2}){\bf k}(n-1){\bf x}^{T}_{N}(n-1){\bf P}(n-2)\right],
\end{eqnarray}
so that \eqref{PPRMCCAlgorithm7} can be rewritten as
\begin{eqnarray}
    \label{PPRMCCAlgorithm81}
    {\bf k}(n)=\lambda^{-1}\bigg[{\bf P}(n-1)-{\rm{exp}}(-\frac{e^2(n|n\!-\!1)}{2\sigma^2}){\bf k}(n){\bf x}^{T}_{N}(n){\bf P}(n-1)\bigg]{\bf x}_{N}(n)={\bf P}(n){\bf x}_{N}(n).
\end{eqnarray}


\subsection{The PRMCC Algorithm}

Motivated by the PRLS from \cite{Qin20}, an $N\times N$ proportionate matrix ${\bf G}(n-1)$ is introduced into the update of \eqref{PPRMCCAlgorithm6}, leading to the PRMCC algorithm within the following update rule:
\begin{eqnarray}
    \label{PPRMCCAlgorithm9}
    {\bf w}_N(n)\!=\!{\bf w}_N(n\!-\!1)+{\bf G}(n\!-\!1){\bf k}(n)f(e(n|n\!-\!1)),
\end{eqnarray}
where ${\bf G}(n-1)=\text{diag}$$\{g_1 (n-1), g_2 (n-1), \ldots, g_N (n-1)\}$ with the $k$-th diagonal element given by
\begin{eqnarray}
    \label{PPRMCCAlgorithm10}
    g_k(n-1)=
    \theta\left[\frac{(1-\alpha)}{2N}+(1+\alpha)\frac{|w_k(n-1)|}{2||{\bf w}_N(n-1)||_1+\epsilon}\right].
\end{eqnarray} 
Here, $\epsilon$ is a small positive constant, $\alpha \in [-1,1)$, and $\theta>0$ is the trace controller. The parameter $\theta$ trades off the convergence and steady-state behaviors of the PRMCC. 
It is noted as $\sigma\to\infty$ in \eqref{PPRMCCAlgorithm61}, the PRMCC algorithm reduces to the PRLS algorithm.

The implementation of PRMCC algorithm is summarized in {\bf{Algorithm}} {\bf{1}}.
\floatname{algorithm}{Algorithm}
\renewcommand{\algorithmicrequire}{\textbf{Initialization:}}
\renewcommand{\algorithmicensure}{\textbf{Iteration:}}
\begin{algorithm}
\begin{algorithmic}
\caption{The PRMCC Algorithm}
\Require ${\bf P}(0)=\delta^{-1}{\bf I}_N$, ${\bf w}_N(0)={\bf 0}_{N\times 1}$, $\lambda$, $\delta$, $\epsilon$, $\theta$, $\alpha$ and $\sigma$ are constants;
\Ensure For $n=1,2,\cdots$ \\
\ $e(n|n-1)=d(n)-{\bf w}^T_{N}(n-1){\bf x}_N(n)$ \\
\ ${\bf k}(n)=\frac{{\bf P}(n-1){\bf x}_{N}(n)}{\lambda+{\rm{exp}}(-\frac{e^2(n|n\!-\!1)}{2\sigma^2}){\bf x}^{T}_{N}(n){\bf P}(n-1){\bf x}_{N}(n)}$ \\
\ $g_k(n-1)=\frac{\theta(1-\alpha)}{2N}+\theta(1+\alpha)\frac{|w_k(n-1)|}{2||{\bf w}_N(n-1)||_1+\epsilon}, k=1,2,\dots N$ \\
\ ${\bf G}(n-1)=\text{diag}\{g_1(n-1),g_2(n-1),...,g_{N}(n-1)\}$ \\
\ $f(e(n|n\!-\!1))={\rm{exp}}(-\frac{e^2(n|n\!-\!1)}{2\sigma^2})e(n|n\!-\!1)$\\
\ ${\bf w}_N(n)={\bf w}_N(n-1)+{\bf G}(n-1){\bf k}(n)f(e(n|n-1))$\\
\ ${\bf P}(n)=\lambda^{-1}[{\bf P}(n\!-\!1)-{\rm{exp}}(-\frac{e^2(n|n\!-\!1)}{2\sigma^2}){\bf k}(n){\bf x}^{T}_{N}(n){\bf P}(n\!-\!1)]$
\end{algorithmic}
\end{algorithm}


\subsection{The CPRMCC Algorithm}

As mentioned before, the convergence and steady-state behaviors of the PRMCC is traded off by the parameter $\theta$ in the proportionate matrix. To simultaneously achieve fast convergence and good steady-state performance, we further propose the CPRMCC algorithm, which is indeed an adaptive convex combination of two PRMCC filters with two different trace controllers. The output of the CPRMCC is given as
\begin{eqnarray}
    \label{CPRMCC1}
    \begin{aligned}
    y(n)&=\rho(n){\bf w}^T_{1}(n){\bf x}_N(n)+(1-\rho(n)){\bf w}^T_{2}(n){\bf x}_N(n)\\
    &={\bf w}_N^T(n){\bf x}_N(n),
    \end{aligned}
\end{eqnarray}
where ${\bf w}_1(n)$ and ${\bf w}_2(n)$ are updated as \eqref{PPRMCCAlgorithm9} with trace controllers being set to $\theta_1$ and $\theta_2$. The overall weight estimate is then 
\begin{eqnarray}
    \label{CPRMCC2}
    {\bf w}_N(n)=\rho(n){\bf w}_1(n)+(1\!-\!\rho(n)){\bf w}_2(n).
\end{eqnarray}
The mixing parameter $\rho(n)$ is defined as a sigmoidal function
\begin{eqnarray}
    \label{CPRMCC3}
    \rho(n)={\rm sgm}[b(n-1)]=\frac{1}{1+{\rm exp}(-b(n-1))},
\end{eqnarray}
where $b(n-1)$ is iteratively updated via \cite{Shi14} 
\begin{align}
    \label{CPRMCC4}
    b(n\!-\!1)=\left[b(n\!-\!2)+\mu_b{\rm{exp}}(-\frac{e^2(n\!-\!1|n\!-\!1)}{2\sigma_b^2})e(n\!-\!1|n\!-\!1)(y_1(n\!-\!1)\!-\!y_2(n\!-\!1))\rho(n\!-\!1)(1-\rho(n\!-\!1))\right]^{b^+}_{-b^+},
\end{align}
in which
\begin{eqnarray}
    \label{CPRMCC111}
    y_1(n)={\bf w}^T_{1}(n){\bf x}_N(n) \ \ \text{and} \ \ \
    y_2(n)={\bf w}^T_{2}(n){\bf x}_N(n),
\end{eqnarray}
and $\mu_b$ and $\sigma_b$ are, respectively, the step size and kernel bandwidth. The values of $b(n-1)$ has been limited to the interval $[-b^+, b^+]$ \cite{Garcia05,Garcia06,Garcia09} to avoid possible pre-mature termination of the update when $\rho(n-1)$ is close to zero or one.

The detailed implementation of CPRMCC algorithm is summarized in {\bf Algorithm} {\bf 2}.
\floatname{algorithm}{Algorithm}
\renewcommand{\algorithmicrequire}{\textbf{Initialization:}}
\renewcommand{\algorithmicensure}{\textbf{Iteration:}}
\begin{algorithm}
\begin{algorithmic}
\caption{The CPRMCC Algorithm}
\Require ${\bf P}_i(0)=\delta^{-1}{\bf I}_N$, ${\bf w}_i(0)={\bf 0}_{N\times 1}$, $i=1,2$, $b(0)=0$, $\lambda$, $\delta$, $\epsilon$, $\alpha$, $\theta_1$, $\theta_2$, $\sigma$, $b^+$, $\sigma_b$ and $\mu_b$ are constants;
\Ensure For $n=1,2,\cdots$ \\
\ $y_i(n)={\bf w}^T_{i}(n-1){\bf x}_N(n), i=1,2$\\
\ $\rho(n)={\rm sgm}(b(n-1))$\\
\ $y(n)=\rho(n)y_1(n)+(1-\rho(n))y_2(n)$\\
\ $e(n|n-1)=d(n)-y(n)$\\
\ $b(n)=[b(n-1)+\mu_b{\rm{exp}}(-\frac{e^2(n|n\!-\!1)}{2\sigma_b^2})e(n|n\!-\!1)(y_1(n)-y_2(n))\rho(n)(1-\rho(n))]^{b^+}_{-b^+}$\\
\ ${\bf w}_i(n), i=1,2$ are updated via the PRMCC algorithm  \\
\ ${\bf w}_N(n)=\rho(n){\bf w}_1(n)+(1\!-\!\rho(n)){\bf w}_2(n)$
\end{algorithmic}
\end{algorithm}


The CPRMCC can be further improved by speeding up the convergence of the slower component PRMCC filter via the faster filter. Assuming ${\bf w}_1(n)$ converges faster than ${\bf w}_2(n)$ thus $\theta_1>\theta_2$, then ${\bf w}_2(n)$ can be accelerated as follows \cite{Shi14}
\begin{eqnarray}
    \label{CPRMCC5}
    {\bf w}_2(n)=\beta({\bf w}_2(n-1)+{\bf G}_2(n-1){\bf k}_2(n)f(e_2(n|n-1)))+(1-\beta){\bf w}_1(n),
\end{eqnarray}
where $0<\beta<1$.

To make it easier for the reader to apply algorithms, the complexity analysis in terms of multiplication per updating (iteration) and hyperparameters are given in Table.~\ref{complexity of different algorithms}, where we omit the complexity of the exponential function.
\begin{table}[!ht]
\renewcommand{\arraystretch}{1.2}
\begin{center}
\caption{The complexity comparison and hyperparameters among different algorithms} 
\label{complexity of different algorithms}
{\begin{tabular}{|c||p{3.5cm}<{\centering}|p{5cm}<{\centering}|}\hline  {\bf Algorithms} & {Complexity in multiplication} & {Hyperparameters}
\\\hline\hline {\bf RLS} & $\mathcal{O}(3N^2+4N)$ & $\lambda$, $\delta$
\\\hline {\bf PRLS} & $\mathcal{O}(3N^2+7N)$ & $\lambda$, $\delta$, $\epsilon$, $\theta$, $\alpha$
\\\hline {\bf PRMCC} & $\mathcal{O}(3N^2+8N )$ & $\lambda$, $\delta$, $\epsilon$, $\theta$, $\alpha$, $\sigma$
\\\hline {\bf CPRMCC} & $\mathcal{O}(6N^2 + 18N )$  & $\lambda$, $\delta$, $\epsilon$, $\theta_1$, $\theta_2$, $\alpha$, $\sigma$, $b^+$, $\mu_b$, $\sigma_b$, $\beta$ \\\hline
\end{tabular}}{}
\end{center}
\end{table}




\section{Performance Analysis}
\label{Perforanalysis}
\vspace{0.4cm}

In this section, performance analysis is made for PRMCC and CPRMCC. First, we analyze the stability condition, steady-state and tracking performance of  PRMCC. After that, we investigate the performance of the CPRMCC. 
Key assumptions used in the analysis are

{\bf Assumption 1}. The noise, ${v(n)}$, is zero-mean, independent and identically distributed (i.i.d.)
and is independent of ${\bf x}_N(m)$ for all $m$ \cite{Ni10,Paul11,Qian19}.

{\bf Assumption 2}. The filter weight vector, ${\bf w}_N(n-1)$, and the regressor, ${\bf x}_N(n)$, are uncorrelated under slow adaption conditions \cite{Haykin14,Das16,QinDSP22}.

{\bf Assumption 3}. The ${\bf x}_N(n)$ is a white Gaussian sequence with variance $\sigma_x^2$ \cite{Sayed09,ChenJ15,Xia17}.

\subsection{Performance Analysis for PRMCC}

\subsubsection{Stability Condition}
\label{Sta}

We first derive the stability condition for PRMCC. Based on \eqref{PPRMCCAlgorithm81}, the weight update of PRMCC in \eqref{PPRMCCAlgorithm9} can be rewritten as
\begin{eqnarray}
    \label{Sta1}
    {\bf w}_N(n)\!=\!{\bf w}_N(n\!-\!1)\!+\!{\bf G}(n\!-\!1){\bf P}(n){\bf x}_{N}(n)f(e(n|n\!\!-\!\!1)).
\end{eqnarray}
Subtracting \eqref{Sta1} by ${\bf w}_N$ on both sides and defining the weight error vector as 
\begin{eqnarray}
    \label{Sta2}
    {\bf h}_N(n)={\bf w}_N-{\bf w}_N(n),
\end{eqnarray}
we have
\begin{eqnarray}
    \label{Sta3}
    {\bf h}_N(n)\!=\!{\bf h}_N(n\!-\!1)\!-\!{\bf G}(n\!-\!1){\bf P}(n){\bf x}_{N}(n)f(e(n|n\!\!-\!\!1)).
\end{eqnarray}
By the detailed proof in {Appendix} \ref{Proof of first order performance of PRMCC}, it can be derived that
\begin{align}
    \label{Sta13}
    &\text{E}[{\bf h}_N(n)]\approx\left[{\bf I}_N\!-\!\frac{(1\!-\!\lambda)(1\!-\!\frac{3\sigma_v^2}{2\sigma^2}\!+\!\frac{5\text{E}[v^4(n)]}{8\sigma^4})}
    {1-\frac{\sigma_v^2}{2\sigma^2}+\frac{\text{E}[v^4(n)]}{8\sigma^4}}{\bf G}\right]^n\text{E}[{\bf h}_N(0)].
\end{align}
where ${\bf G}$ is defined in \eqref{Sta8}.


As ${\bf I}_N-\frac{(1-\lambda)(1-\frac{3\sigma_v^2}{2\sigma^2}+\frac{5\text{E}[v^4(n)]}{8\sigma^4})}{1-\frac{\sigma_v^2}{2\sigma^2}
+\frac{\text{E}[v^4(n)]}{8\sigma^4}}{\bf G}$ is diagonal, the condition for the right-hand side (RHS) of \eqref{Sta13} to converge to zero as $n$ $\rightarrow$ $\infty$ is that all diagonal elements should have magnitude less than one. Equivalently, we require
\begin{align}
    \label{Sta14}
    0<\theta<\frac{2(1-\frac{\sigma_v^2}{2\sigma^2}+\frac{\text{E}[v^4(n)]}{8\sigma^4})}{(1\!-\!\lambda)(1\!-\!\frac{3\sigma_v^2}{2\sigma^2}
\!+\!\frac{5\text{E}[v^4(n)]}{8\sigma^4})(\frac{1-\alpha}{2N}\!+\!\frac{(1+\alpha)||{\bf w}_N||_{\infty}}{2||{\bf w}_N||_1})},
\end{align}
which is the stability condition of the PRMCC algorithm. When $\sigma\rightarrow\infty$ in particular, the condition reduces to that of the PRLS \cite{QinDSP22}
\begin{align}
    \label{Sta15}
    0<\theta<\frac{2}{(1\!-\!\lambda)(\frac{1-\alpha}{2N}\!+\!\frac{(1+\alpha)||{\bf w}_N||_{\infty}}{2||{\bf w}_N||_1})}.
\end{align}

We continue to take the $l_2$ norm on both sides of \eqref{Sta13} and obtain
\begin{align}
    \label{Sta16}
    ||\text{E}[{\bf h}_N(n)]||_2&\approx||[{\bf I}_N\!-\!\frac{(1\!-\!\lambda)(1\!-\!\frac{3\sigma_v^2}{2\sigma^2}\!+\!\frac{5\text{E}[v^4(n)]}{8\sigma^4})}
    {1-\frac{\sigma_v^2}{2\sigma^2}+\frac{\text{E}[v^4(n)]}{8\sigma^4}}{\bf G}]^n\text{E}[{\bf h}_N(0)]||_2\nonumber\\
    &\leq \bigg(1  - \frac{\theta(1\!-\!\lambda)(1\!-\!\frac{3\sigma_v^2}{2\sigma^2}\!+\!\frac{5\text{E}[v^4(n)]}{8\sigma^4})(\frac{1-\alpha}{2N}+\frac{(1+\alpha)|w_{\min}|}{2||{\bf w}_N||_1})}
    {1-\frac{\sigma_v^2}{2\sigma^2}+\frac{\text{E}[v^4(n)]}{8\sigma^4}}  \bigg)^n||\text{E}[{\bf h}_N(0)]||_2,
\end{align}
where $|w_{\min}|$ denotes the smallest absolute value in ${\bf w}_N$. Based on \eqref{Sta16},  the maximum number of iterations for the PRMCC to reach a steady-state error $c$ is
\begin{align}
    \label{Sta17}
    N_m= \left\lceil\frac{{\text{ln}}c-{\text{ln}}||\text{E}[{\bf h}_N(0)]||_2  }{{\text{ln}}\bigg( 1  - \frac{\theta(1\!-\!\lambda)(1\!-\!\frac{3\sigma_v^2}{2\sigma^2}\!+\!\frac{5\text{E}[v^4(n)]}{8\sigma^4})(\frac{1-\alpha}{2N}+\frac{(1+\alpha)|w_{\min}|}{2||{\bf w}_N||_1})}
    {1-\frac{\sigma_v^2}{2\sigma^2}+\frac{\text{E}[v^4(n)]}{8\sigma^4}} \bigg)}\right\rceil,
\end{align}
where $c\leq ||\text{E}[{\bf h}_N(0)]||_2$.
As expected, the larger is the trace controller parameter $\theta$ of ${\bf G}$ is, the less number of iterations is needed for PRMCC to reach the steady state.

\subsubsection{Steady-state Performance Analysis}
\label{SSA}

In this part, we derive the steady-state mean-square deviation (MSD) for the PRMCC. Computing the autocorrelations on both sides of \eqref{Sta3}, one obtains
\begin{eqnarray}
    \label{SSA1}
    {\bf K}(n)&\!\!\!=\!\!\!&{\bf K}(n-1)-\text{E}[{\bf h}_N(n\!-\!1){\bf x}^T_{N}(n){\bf P}(n){\bf G}(n\!-\!1) f(e(n|n\!-\!1))]-\text{E}[{\bf G}(n\!-\!1){\bf P}(n){\bf x}_{N}(n){\bf h}^T_N(n\!-\!1)f(e(n|n\!-\!1))]\nonumber\\
    &&+\text{E}[{\bf G}(n\!-\!1){\bf P}(n){\bf x}_{N}(n){\bf x}^T_{N}(n){\bf P}(n){\bf G}(n\!-\!1)f^2(e(n|n\!-\!1))],
\end{eqnarray}
where ${\bf K}(n)=\text{E}[{\bf h}_N(n){\bf h}^T_N(n)]$. Based on the derivations in {Appendix} \ref{proof of second order mean square performance of PRMCC}, we have
\begin{align}
    \label{SSA7}
    {\bf K}_i(\infty) \approx \frac{\frac{(1-\lambda)g_i(\sigma_v^2-\frac{\text{E}[v^4(n)]}{\sigma^2}+\frac{\text{E}[v^6(n)]}{2\sigma^4})}
    {\sigma_x^2(1-\frac{\sigma_v^2}{2\sigma^2}+\frac{\text{E}[v^4(n)]}{8\sigma^4})}}{2\!-\!\frac{3\sigma_v^2}{\sigma^2}
    \!+\!\frac{5\text{E}[v^4(n)]}{4\sigma^4}-\frac{(1-\lambda)g_i(1-\frac{6\sigma_v^2}{\sigma^2}+\frac{15\text{E}[v^4(n)]}{2\sigma^4})}
    {1-\frac{\sigma_v^2}{2\sigma^2}+\frac{\text{E}[v^4(n)]}{8\sigma^4}}},
\end{align}
where ${\bf K}_i(\infty)$ is the $(i,i)$-th element of ${\bf K}(\infty)$. The steady-state ${\rm MSD}$ of the PRMCC is then obtained as
\begin{align}
    \label{SSA8}
    {\rm MSD}_{\rm PRMCC}=\sum_{i=1}^{N}{\bf K}_i(\infty)\approx \sum_{i=1}^{N}\frac{\frac{(1-\lambda)g_i(\sigma_v^2-\frac{\text{E}[v^4(n)]}{\sigma^2}+\frac{\text{E}[v^6(n)]}{2\sigma^4})}
    {\sigma_x^2(1-\frac{\sigma_v^2}{2\sigma^2}+\frac{\text{E}[v^4(n)]}{8\sigma^4})}}{2\!-\!\frac{3\sigma_v^2}{\sigma^2}
    \!+\!\frac{5\text{E}[v^4(n)]}{4\sigma^4}-\frac{(1-\lambda)g_i(1-\frac{6\sigma_v^2}{\sigma^2}+\frac{15\text{E}[v^4(n)]}{2\sigma^4})}
    {1-\frac{\sigma_v^2}{2\sigma^2}+\frac{\text{E}[v^4(n)]}{8\sigma^4}}}.
\end{align}
It is noted that when ${\bf G}(n)={\bf I}_N$, the PRMCC reduces to the RMCC and we have
\begin{align}
    \label{SSA9}
    &{\rm MSD}_{\rm RMCC}\approx\frac{\frac{N(1-\lambda)(\sigma_v^2-\frac{\text{E}[v^4(n)]}{\sigma^2}+\frac{\text{E}[v^6(n)]}{2\sigma^4})}
    {\sigma_x^2(1-\frac{\sigma_v^2}{2\sigma^2}+\frac{\text{E}[v^4(n)]}{8\sigma^4})}}{2\!-\!\frac{3\sigma_v^2}{\sigma^2}
    \!+\!\frac{5\text{E}[v^4(n)]}{4\sigma^4}-\frac{(1-\lambda)(1-\frac{6\sigma_v^2}{\sigma^2}+\frac{15\text{E}[v^4(n)]}{2\sigma^4})}
    {1-\frac{\sigma_v^2}{2\sigma^2}+\frac{\text{E}[v^4(n)]}{8\sigma^4}}}.
\end{align}
Moreover, if $\sigma\to\infty$, the MSD of PRLS and RLS can also be conveniently derived from \eqref{SSA9} as
\begin{align}
    \label{SSA10}
    &{\rm MSD}_{\rm PRLS}\approx \sum_{i=1}^{N}\frac{(1-\lambda)g_i\sigma_v^2}{2\sigma_x^2-(1-\lambda)g_i\sigma_x^2},
\end{align}

\begin{align}
    \label{SSA11}
    &{\rm MSD}_{\rm RLS}\approx\frac{N(1-\lambda)\sigma_v^2}{2\sigma_x^2-(1-\lambda)\sigma_x^2}.
\end{align}

\subsubsection{Tracking Performance Analysis}
\label{TA}

In this part, we proceed to evaluate the tracking ability of PRMCC in nonstationary environments. Without loss of generality, it is assumed that the unknown time-varying system vector ${\bf w}(n)$ follows a random-walk model \cite{Sayed09}
\begin{eqnarray}
    \label{TA1}
    {\bf w}(n)={\bf w}(n-1)+{\bf q}(n),
\end{eqnarray}
where ${\bf q}(n)$ is an i.i.d. white Gaussian process noise with zero mean and covariance matrix $E[{\bf q}(n){\bf q}^T(n)]=\sigma_q^2 {\bf I}_N$, and ${\bf w}(0)={\bf w}_N$.

We redefine the weight error vector in \eqref{Sta2} as ${\bf h}_N(n)={\bf w}(n)-{\bf w}_N(n)$. Subtracting both sides of \eqref{Sta1} by ${\bf w}(n)$ leads to
\begin{eqnarray}
    \label{TA2}
    {\bf h}_N(n)&\!\!=\!\!&{\bf h}_N(n-1)-{\bf G}(n-1){\bf P}(n){\bf x}_{N}(n) f(e(n|n-1))+{\bf q}(n).
\end{eqnarray}
Computing the autocorrelations on both sides of \eqref{TA2}, one obtains
\begin{eqnarray}
    \label{TA3}
    {\bf K}(n)&\!\!\!=\!\!\!&{\bf K}(n-1)-\text{E}[{\bf h}_N(n\!-\!1){\bf x}^T_{N}(n){\bf P}(n){\bf G}(n\!-\!1) f(e(n|n\!-\!1))]-\text{E}[{\bf G}(n\!-\!1){\bf P}(n){\bf x}_{N}(n){\bf h}^T_N(n\!-\!1)f(e(n|n\!-\!1))]\nonumber\\
    &&+\text{E}[{\bf G}(n\!-\!1){\bf P}(n){\bf x}_{N}(n){\bf x}^T_{N}(n){\bf P}(n){\bf G}(n\!-\!1)f^2(e(n|n\!-\!1))]+\sigma_q^2{\bf I}_N.
\end{eqnarray}
Plugging \eqref{SSA2}, \eqref{SSA3} and \eqref{SSA5} into \eqref{TA3}, we obtain the steady-state MSD in nonstationary environments for the PRMCC as 
\begin{align}
    \label{TA4}
    \widetilde{\rm MSD}_{\rm PRMCC}\approx \sum_{i=1}^{N}\frac{\frac{(1-\lambda)g_i(\sigma_v^2-\frac{\text{E}[v^4(n)]}{\sigma^2}+\frac{\text{E}[v^6(n)]}{2\sigma^4})}
    {\sigma_x^2(1-\frac{\sigma_v^2}{2\sigma^2}+\frac{\text{E}[v^4(n)]}{8\sigma^4})}+\frac{\sigma_q^2(1-\frac{\sigma_v^2}{2\sigma^2}+
    \frac{\text{E}[v^4(n)]}{8\sigma^4})}{(1-\lambda)g_i}}{2\!-\!\frac{3\sigma_v^2}{\sigma^2}
    \!+\!\frac{5\text{E}[v^4(n)]}{4\sigma^4}-\frac{(1-\lambda)g_i(1-\frac{6\sigma_v^2}{\sigma^2}+\frac{15\text{E}[v^4(n)]}{2\sigma^4})}
    {1-\frac{\sigma_v^2}{2\sigma^2}+\frac{\text{E}[v^4(n)]}{8\sigma^4}}}.
\end{align}
As $2\!-\!\frac{3\sigma_v^2}{\sigma^2}\!+\!\frac{5\text{E}[v^4(n)]}{4\sigma^4}\gg\frac{(1-\lambda)g_i(1-\frac{6\sigma_v^2}
{\sigma^2}+\frac{15\text{E}[v^4(n)]}{2\sigma^4})}{1-\frac{\sigma_v^2}{2\sigma^2}+\frac{\text{E}[v^4(n)]}{8\sigma^4}}$, \eqref{TA4} can be further simplified into
\begin{align}
    \label{TA5}
    \widetilde{\rm MSD}_{\rm PRMCC}\approx \frac{\frac{(1-\lambda)\theta(\sigma_v^2-\frac{\text{E}[v^4(n)]}{\sigma^2}+\frac{\text{E}[v^6(n)]}{2\sigma^4})}
    {\sigma_x^2(1-\frac{\sigma_v^2}{2\sigma^2}+\frac{\text{E}[v^4(n)]}{8\sigma^4})}\!+\!\frac{\sigma_q^2(1-\frac{\sigma_v^2}{2\sigma^2}+
    \frac{\text{E}[v^4(n)]}{8\sigma^4})\sum_{i=1}^{N}{t_i^{-1}}}{(1-\lambda)\theta}}{2\!-\!\frac{3\sigma_v^2}{\sigma^2}
    \!+\!\frac{5\text{E}[v^4(n)]}{4\sigma^4}},
\end{align}
where $t_i=\frac{1-\alpha}{2N}+\frac{(1+\alpha)|w_i|}{2||{\bf w}_N||_1}$.

A close observation reveals that the $\widetilde{\rm MSD}_{\rm PRMCC}$ in \eqref{TA5} is a function of $\theta$ and $\lambda$.
By taking the partial derivatives and setting the results to zeros, i.e., $\frac{\partial{\widetilde{\rm MSD}_{\rm PRMCC}}}{\partial{\theta}}=0$ and $\frac{\partial{\widetilde{\rm MSD}_{\rm PRMCC}}}{\partial{\lambda}}=0$, we can obtain the theoretically optimal parameters as 
\begin{eqnarray}
    \label{TA6}
    \theta_{\rm opt}=\frac{\sigma_x\sigma_q(1-\frac{\sigma_v^2}{2\sigma^2}+\frac{\text{E}[v^4(n)]}{8\sigma^4})}
    {(1-\lambda)\sqrt{\sigma_v^2-\frac{\text{E}[v^4(n)]}{\sigma^2}+\frac{\text{E}[v^6(n)]}{2\sigma^4}}}\sqrt{\sum_{i=1}^{N}{t_i^{-1}}},
\end{eqnarray}
\begin{eqnarray}
    \label{TA7}
    \lambda_{\rm opt}=1-\frac{\sigma_x\sigma_q(1-\frac{\sigma_v^2}{2\sigma^2}+\frac{\text{E}[v^4(n)]}{8\sigma^4})}
    {\theta\sqrt{\sigma_v^2-\frac{\text{E}[v^4(n)]}{\sigma^2}+\frac{\text{E}[v^6(n)]}{2\sigma^4}}}\sqrt{\sum_{i=1}^{N}{t_i^{-1}}}.
\end{eqnarray}

\subsection{Performance Analysis for the CPRMCC}
\label{CPRMCC}

\subsubsection{Mean-square Performance Analysis}

According to \eqref{CPRMCC2} and \eqref{Sta2}, we have
\begin{eqnarray}
    \label{CPRMCC6}
    {\bf h}_N(n)=\rho(n){\bf h}_1(n)+(1-\rho(n)){\bf h}_2(n),
\end{eqnarray}
where ${\bf h}_1(n)={\bf w}_N-{\bf w}_1(n)$ and ${\bf h}_2(n)={\bf w}_N-{\bf w}_2(n)$ are weight errors of the two PRMCC filters in CPRMCC.
Computing the autocorrelations on both sides of \eqref{CPRMCC6}, one obtains
\begin{eqnarray}
    \label{CPRMCC7}
    {\bf K}(n)&\!\!\!\!=\!\!\!\!&{\rm E}[\rho^2(n){\bf h}_1(n){\bf h}_1^T(n)]\!\!+\!\!{\rm E}[(1-\rho(n))^2{\bf h}_2(n){\bf h}_2^T(n)]+{\rm E}[\rho(n)(1-\rho(n))({\bf h}_1(n){\bf h}_2^T(n)+{\bf h}_2(n){\bf h}_1^T(n))]\nonumber\\
    &\!\!\!\!\approx\!\!\!\!&{\rm E}[\rho^2(n)]{\bf H}_{11}(n)+{\rm E}[(1-\rho(n))^2]{\bf H}_{22}(n)+{\rm E}[\rho(n)(1-\rho(n))]({\bf H}_{12}(n)+{\bf H}_{21}(n)),
\end{eqnarray}
where ${\bf H}_{ij}(n)={\rm E}[{\bf h}_i(n){\bf h}_j^T(n)]$, $i,j=1,2$.
The approximation in the second line is valid because $\rho(n)$ is independent of both component filters when $\rho(n)(1-\rho(n))$ is close to zero in the steady state \cite{Garcia06}.

Taking the trace of \eqref{CPRMCC7} and letting $n$$\to$$\infty$, the steady-state MSD of the proposed CPRMCC algorithm is
\begin{eqnarray}
    \label{CPRMCC8}
    {\rm MSD}_{\rm CPRMCC}\approx\lim\limits_{n\to\infty}\text{E}[\rho^2(n)]{\rm MSD}_1\!\!+\!\!\lim\limits_{n\to\infty}\text{E}[(1-\rho(n))^2]{\rm MSD}_2+2\lim\limits_{n\to\infty}\text{E}[\rho(n)(1-\rho(n))]{\rm MSD}_{12},
\end{eqnarray}
where 
\begin{eqnarray}
    \label{CPRMCC81}
    {\rm MSD}_j={\rm tr}({\bf H}_{jj}(\infty)), \ j=1,2,
\end{eqnarray}
\begin{eqnarray}
    \label{CPRMCC82}
    {\rm MSD}_{12}={\rm tr}({\bf H}_{12}(\infty))={\rm tr}({\bf H}_{21}(\infty)).
\end{eqnarray}
According to \eqref{SSA8}, we have
\begin{eqnarray}
    \label{CPRMCC9}
    {\rm MSD}_{j}\approx \sum_{i=1}^{N}\frac{\frac{(1-\lambda)\theta_j t_i(\sigma_v^2-\frac{\text{E}[v^4(n)]}{\sigma^2}+\frac{\text{E}[v^6(n)]}{2\sigma^4})}
    {\sigma_x^2(1-\frac{\sigma_v^2}{2\sigma^2}+\frac{\text{E}[v^4(n)]}{8\sigma^4})}}{2\!-\!\frac{3\sigma_v^2}{\sigma^2}
    \!+\!\frac{5\text{E}[v^4(n)]}{4\sigma^4}-\frac{(1-\lambda)\theta_j t_i(1-\frac{6\sigma_v^2}{\sigma^2}+\frac{15\text{E}[v^4(n)]}{2\sigma^4})}
    {1-\frac{\sigma_v^2}{2\sigma^2}+\frac{\text{E}[v^4(n)]}{8\sigma^4}}}, \nonumber\\
\end{eqnarray} 
Based on the derivation in {Appendix} \ref{Proof of combination of two PRMCC in the steady state}
\begin{align}
    \label{CPRMCC18}
    {\rm MSD}_{12}\approx \sum_{i=1}^{N}\frac{\frac{(1-\lambda)\theta_{12}t_i(\sigma_v^2-\frac{\text{E}[v^4(n)]}{\sigma^2}+\frac{\text{E}[v^6(n)]}{4\sigma^4})}
    {\sigma_x^2(1-\frac{\sigma_v^2}{2\sigma^2}+\frac{\text{E}[v^4(n)]}{8\sigma^4})}}{(1\!-\!\frac{3\sigma_v^2}{2\sigma^2}
    \!+\!\frac{5\text{E}[v^4(n)]}{8\sigma^4})-\frac{(1-\lambda)\theta_{12}t_i(1-\frac{3\sigma_v^2}{\sigma^2}+\frac{9\text{E}[v^4(n)]}{4\sigma^4})}
    {1-\frac{\sigma_v^2}{2\sigma^2}+\frac{\text{E}[v^4(n)]}{8\sigma^4}}},
\end{align}
with
\begin{eqnarray}
    \label{CPRMCC19}
    \theta_{12}=\frac{\theta_1\theta_2}{\theta_1+\theta_2}.
\end{eqnarray}

Last, by resorting to \eqref{TA4}, the steady-state ${\rm MSD}_{j}$, $j=1,2$ and ${\rm MSD}_{12}$ in the nonstationary environments can also be derived
\begin{align}
    \label{CPRMCC20_1}
    \widetilde{{\rm MSD}}_{j}\approx \sum_{i=1}^{N}\frac{\frac{(1-\lambda)\theta_j t_i(\sigma_v^2-\frac{\text{E}[v^4(n)]}{\sigma^2}+\frac{\text{E}[v^6(n)]}{2\sigma^4})}
    {\sigma_x^2(1-\frac{\sigma_v^2}{2\sigma^2}+\frac{\text{E}[v^4(n)]}{8\sigma^4})}\!+\!\frac{\sigma_q^2(1-\frac{\sigma_v^2}{2\sigma^2}+
    \frac{\text{E}[v^4(n)]}{8\sigma^4})}{(1-\lambda)\theta_j t_i}}{2\!-\!\frac{3\sigma_v^2}{\sigma^2}
    \!+\!\frac{5\text{E}[v^4(n)]}{4\sigma^4}\!-\!\frac{(1-\lambda)\theta_j t_i(1-\frac{6\sigma_v^2}{\sigma^2}+\frac{15\text{E}[v^4(n)]}{2\sigma^4})}
    {1-\frac{\sigma_v^2}{2\sigma^2}+\frac{\text{E}[v^4(n)]}{8\sigma^4}}},
\end{align}
\begin{align}
    \label{CPRMCC20}
    \widetilde{{\rm MSD}}_{12}\approx \sum_{i=1}^{N}\frac{\frac{(1-\lambda)\theta_{12}t_i(\sigma_v^2-\frac{\text{E}[v^4(n)]}{\sigma^2}+\frac{\text{E}[v^6(n)]}{4\sigma^4})}
    {\sigma_x^2(1-\frac{\sigma_v^2}{2\sigma^2}+\frac{\text{E}[v^4(n)]}{8\sigma^4})}+\frac{\sigma_q^2(1-\frac{\sigma_v^2}{2\sigma^2}+
    \frac{\text{E}[v^4(n)]}{8\sigma^4})}{(1-\lambda)t_i}}{(1\!-\!\frac{3\sigma_v^2}{2\sigma^2}
    \!+\!\frac{5\text{E}[v^4(n)]}{8\sigma^4})-\frac{(1-\lambda)\theta_{12}t_i(1-\frac{3\sigma_v^2}{\sigma^2}+\frac{9\text{E}[v^4(n)]}{4\sigma^4})}
    {1-\frac{\sigma_v^2}{2\sigma^2}+\frac{\text{E}[v^4(n)]}{8\sigma^4}}}.
\end{align}

\subsubsection{Steady-State Behavior of the Mixing Parameter}

As shown in above, the mixing parameter $\rho(n)$ of CPRMCC is a function of $b(n)$, which is iteratively calculated as in \eqref{CPRMCC4}.
Taking expectations on both sides of \eqref{CPRMCC4}, we arrive at
\begin{eqnarray}
    \label{CPRMCC21}
    \text{E}[b(n)]\approx\bigg[\text{E}[b(n\!-\!1)]+\mu_b\text{E}[{\rm{exp}}(-\frac{e^2(n|n\!-\!1)}{2\sigma_b^2})e(n|n\!-\!1)(y_1(n)\!-\!y_2(n))\rho(n)(1-\rho(n))]\bigg]^{b^+}_{-b^+},
\end{eqnarray}
where the right-hand side is an approximation as the order of expectation and truncation operations have been switched in the evaluation.

Based on \eqref{CPRMCC6}, \eqref{Sta9}, {\bf Assumptions 1 {\rm and} 3}, we have
\begin{eqnarray}
    \label{CPRMCC22}
    &&\text{E}[b(n)]\nonumber\\
    &\!\!\!\approx\!\!\!&[\text{E}[b(n\!-\!1)]+\mu_b\text{E}[(1\!-\!\frac{3v^2(n)}{2\sigma^2}\!+\!\frac{5v^4(n)}{8\sigma^4})(\rho(n){\bf h}_1^T(n\!-\!1){\bf x}_{N}(n)+(1\!-\!\rho(n)){\bf h}_2^T(n\!-\!1){\bf x}_{N}(n))({\bf x}_{N}^T(n){\bf h}_2(n-1)\nonumber\\
    &&-{\bf x}_{N}^T(n){\bf h}_1(n\!-\!1))\rho(n)(1-\rho(n))\!+\!(v(n)\!-\!\frac{v^3(n)}{2\sigma^2}\!+\!\frac{v^5(n)}{8\sigma^4})({\bf x}_{N}^T(n){\bf h}_2(n\!-\!1)-{\bf x}_{N}^T(n){\bf h}_1(n\!-\!1))\rho(n)(1\!-\!\rho(n))]]^{b^+}_{-b^+}\nonumber\\
    &\!\!\!\approx\!\!\!&[\text{E}[b(n\!-\!1)]+\mu_b\sigma_x^2(1\!-\!\frac{3\sigma_v^2}{2\sigma^2}\!+\!\frac{5\text{E}[v^4(n)]}{8\sigma^4})\text{E}[\rho^2(n)(1\!-\!\rho(n))({\bf h}_2(n\!-\!1){\bf h}_1^T(n\!-\!1)\nonumber\\
    &&-{\bf h}_1(n\!-\!1){\bf h}_1^T(n\!-\!1))+\rho(n)(1\!-\!\rho(n))^2({\bf h}_2(n\!-\!1){\bf h}_2^T(n\!-\!1)-{\bf h}_1(n\!-\!1){\bf h}_2^T(n\!-\!1))]]^{b^+}_{-b^+}.
\end{eqnarray}
Using the assumption of \eqref{CPRMCC7}, \eqref{CPRMCC22} further becomes
 \begin{align}
    \label{CPRMCC23}
    \text{E}[b(n)]\!\approx\![\text{E}[b(n\!-\!1)]+\mu_b\sigma_x^2(1\!-\!\frac{3\sigma_v^2}{2\sigma^2}\!+\!\frac{5\text{E}[v^4(n)]}{8\sigma^4})(\text{E}[\rho(n)(1\!\!-\!\!\rho(n))^2]\Delta{\rm MSD}_2\!\!-\!\!\text{E}[\rho^2(n)(1\!\!-\!\!\rho(n))]\Delta{\rm MSD}_1)]^{b^+}_{-b^+},
\end{align}
where
\begin{eqnarray}
    \label{CPRMCC24}
    \Delta{\rm MSD}_i={\rm MSD}_i-{\rm MSD}_{12}, i=1,2.
\end{eqnarray}

From \eqref{CPRMCC23}, it is obvious that $\text{E}[b(n)]$ depends on the values of $\Delta{\rm MSD}_i, i=1,2$. According to the Cauchy-Schwartz inequality, ${\rm MSD}_{12}$ cannot be larger than ${\rm MSD}_1$ and ${\rm MSD}_2$ \cite{Garcia06}. Therefore, we only need to discuss the following three cases:
\begin{itemize}
  \item {\bf Case 1}: ${\rm MSD}_1\le{\rm MSD}_{12}\le{\rm MSD}_2$. In this case, $\Delta{\rm MSD}_1\le0$ and $\Delta{\rm MSD}_2\ge0$.
  In addition, since $\rho(n)\in[1-\rho^+,\rho^+]$, both $\text{E}[\rho(n)(1\!\!-\!\!\rho(n))^2]$ and $\text{E}[\rho^2(n)(1\!\!-\!\!\rho(n))]$ are lower bounded by $\rho^+(1\!\!-\!\!\rho^+)^2$.
  As a result, when $n$$\to$$\infty$
  \begin{eqnarray}
    \label{CPRMCC25}
    \text{E}[b(n)]\ge[\text{E}[b(n-1)]+\mu_b\sigma_x^2(1-\frac{3\sigma_v^2}{2\sigma^2}+\frac{5\text{E}[v^4(n)]}{8\sigma^4})\rho^+(1-\rho^+)^2(\Delta{\rm MSD}_2-\Delta{\rm MSD}_1)]^{b^+}_{-b^+}.
  \end{eqnarray}
Therefore, it follows that the only valid stationary point for \eqref{CPRMCC23} is $b(\infty)=b^+$ and the MSD in \eqref{CPRMCC8} then becomes ${\rm MSD}_{\rm CPRMCC}\approx{\rm MSD}_1$.
  \item {\bf Case 2}: ${\rm MSD}_2\le{\rm MSD}_{12}\le{\rm MSD}_1$. In this case, $\Delta{\rm MSD}_1\ge0$ and $\Delta{\rm MSD}_2\le0$, and a similar analysis to the first case can be made.
  As $n$$\to$$\infty$, it is easy to have
  \begin{eqnarray}
    \label{CPRMCC26}
    \text{E}[b(n)]\le[\text{E}[b(n-1)]-\mu_b\sigma_x^2(1-\frac{3\sigma_v^2}{2\sigma^2}+\frac{5\text{E}[v^4(n)]}{8\sigma^4})(\rho^+)^2(1-\rho^+)(\Delta{\rm MSD}_1-\Delta{\rm MSD}_2)]^{b^+}_{-b^+}.
  \end{eqnarray}
Further, $b(\infty)=-b^+$ and the MSD in \eqref{CPRMCC8} is evaluated as ${\rm MSD}_{\rm CPRMCC}\approx{\rm MSD}_2$.
  \item {\bf Case 3}: ${\rm MSD}_{12}<{\rm MSD}_i$, $i=1,2$. In this case, $\Delta{\rm MSD}_i>0$, $i=1,2$.
  A stationary point of \eqref{CPRMCC23} can be represented as
  \begin{eqnarray}
    \label{CPRMCC27}
    \text{E}[\rho(n)(1-\rho(n))^2]\Delta{\rm MSD}_2=\text{E}[\rho^2(n)(1-\rho(n))]\Delta{\rm MSD}_1, \ \ n\to\infty.
  \end{eqnarray}
  With the assumption that the variance of $\rho(n)$ is small when $n$$\to$$\infty$ as in \cite{Garcia06}, \eqref{CPRMCC27} can be rewritten as
  \begin{eqnarray}
    \label{CPRMCC28}
  (1-{\rho}(\infty))\Delta{\rm MSD}_2={\rho}(\infty)\Delta{\rm MSD}_1,
  \end{eqnarray}
  where ${\rho(\infty)}=\lim\limits_{n\to\infty}\text{E}[\rho(n)]$. From \eqref{CPRMCC28}, ${\rho(\infty)}$ is solved as
  \begin{eqnarray}
    \label{CPRMCC29}
  {\rho(\infty)}=[\frac{\Delta{\rm MSD}_2}{\Delta{\rm MSD}_1+\Delta{\rm MSD}_2}]^{b^+}_{-b^+}.
  \end{eqnarray}
  Now, \eqref{CPRMCC8} can be rewritten as
  \begin{eqnarray}
    \label{CPRMCC291}
    {\rm MSD}_{\rm CPRMCC}\approx\rho^2(\infty){\rm MSD_1}\!\!+\!\!(1-\rho(\infty))^2{\rm MSD_2}+2\rho(\infty)(1-\rho(\infty)){\rm MSD_{12}}.
\end{eqnarray}
  Plugging \eqref{CPRMCC29} without the truncation into \eqref{CPRMCC291}, leads to
  \begin{eqnarray}
    \label{CPRMCC30}
    {\rm MSD}_{\rm CPRMCC}\approx{\rm MSD_{12}}+\frac{\Delta{\rm MSD}_1\Delta{\rm MSD}_2}{\Delta{\rm MSD}_1+\Delta{\rm MSD}_2}.
  \end{eqnarray}
  So when $1-\rho^+<\rho(\infty)<\rho^+<1$, based on \eqref{CPRMCC24} and \eqref{CPRMCC29} we have the following approximations:
  \begin{eqnarray}
    \label{CPRMCC31}
    &&\hspace{-1.5cm}{\rm MSD}_{\rm CPRMCC}\approx{\rm MSD_{12}}+\rho(\infty)\Delta{\rm MSD}_1<{\rm MSD}_1,\\
    \label{CPRMCC32}
    &&\hspace{-1.5cm}{\rm MSD}_{\rm CPRMCC}\approx{\rm MSD_{12}}+(1-\rho(\infty))\Delta{\rm MSD}_2<{\rm MSD}_2,
  \end{eqnarray}
  from which we obtain when $-b^+<b(\infty)<b^+$ the following
  \begin{eqnarray}
    \label{CPRMCC33}
    {\rm MSD}_{\rm CPRMCC}<{\rm min}\{{\rm MSD}_1,{\rm MSD}_2\}.
  \end{eqnarray}
\end{itemize}

In summary, we have
\begin{align}
    \label{CPRMCC34}
    \begin{split}
    {\rm MSD}_{\rm CPRMCC} \left \{
        \begin{array}{ll}
        \approx{\rm MSD}_1,                    & b(\infty)=b^+\\
        <{\rm min}\{{\rm MSD}_1,{\rm MSD}_2\},     & -b^+<b(\infty)<b^+\\
        \approx{\rm MSD}_2,                                 & b(\infty)=-b^+
        \end{array}
    \right.
    \end{split},
\end{align}
which indicates the steady-state performance of CPRMCC algorithm is at least as good as that of the better component PRMCC filter.

\section{Simulation Results}
\label{SIMRESULTS}

To verify the performance of PRMCC and CPRMCC algorithms, numerical experiments of system identification were conducted.

\begin{figure}[!ht] 
   \hspace{-0.4cm}%
   \centering
   \includegraphics[width=12cm,keepaspectratio]
   {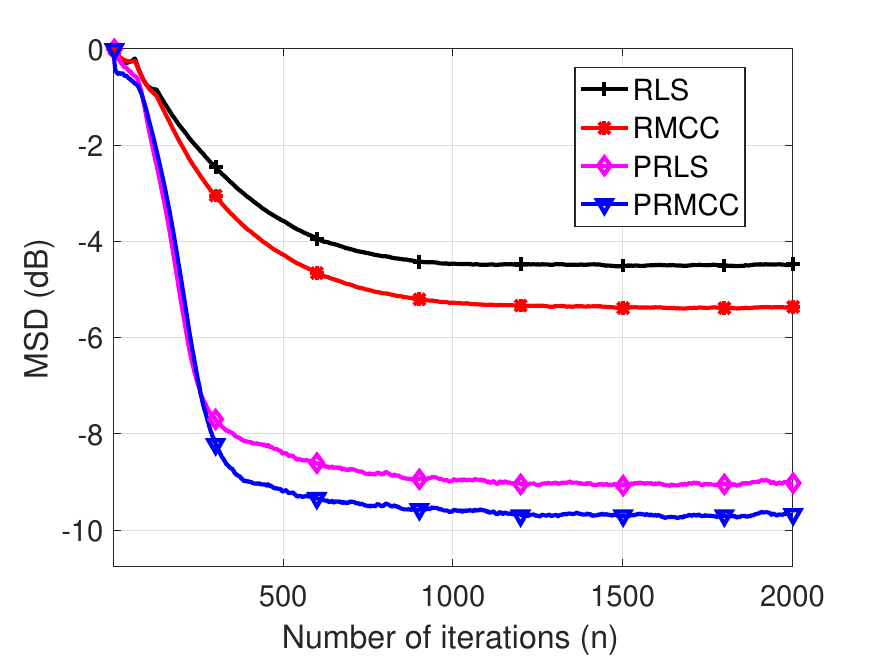}
   \caption{The MSD performance comparison between PRMCC and existing algorithms.}
   \label{Figure_summary1}
\end{figure}

In the first experiment, we aim to show the superiority of the PRMCC by comparing with existing adaptive filtering algorithms including RLS, RMCC and PRLS. A sparse system of order 128 was considered. Its tap values are 0.4975 at indices 1, 64, 65, 128, and 0.0498 at indices 2, 63, 66, 127. At other indices, the values are zeros.
The input signal, $x(n)$ was a zero-mean white Gaussian process with variance $\sigma_x^2=1$.
The observation noise $v(n)$ followed a mixed-Gaussian distribution \cite{Chen17}, that is $v(n)\sim0.9\mathcal{N}(0,10^{-4})+0.1\mathcal{N}(0,100)$. 
The common parameters for RLS, RMCC, PRLS and PRMCC were $\lambda=0.995$ and $\delta=100$.
Specific parameters required to implement PRLS and PRMCC were set as $\theta=64$, $\alpha=0$ and $\epsilon=10^{-4}$. For RMCC and PRMCC, the kernel bandwidth was set as $\sigma=1.7$.
All the simulation results were obtained by averaging over 5000 independent trials.
In Fig.~\ref{Figure_summary1}, MSDs of different algorithms are compared. From the figure, the PRMCC significantly outperforms the RMCC in terms of convergence and steady-state error, attributed to the introduction of the proportionate matrix as shown in \eqref{PPRMCCAlgorithm9}. The RMCC-type algorithms have better performance than their RLS-type correspondences under heavily-tailed noises. 


%
\begin{figure}[!ht] 
   \hspace{-0.4cm}%
   \centering
   \includegraphics[width=12cm,keepaspectratio]
   {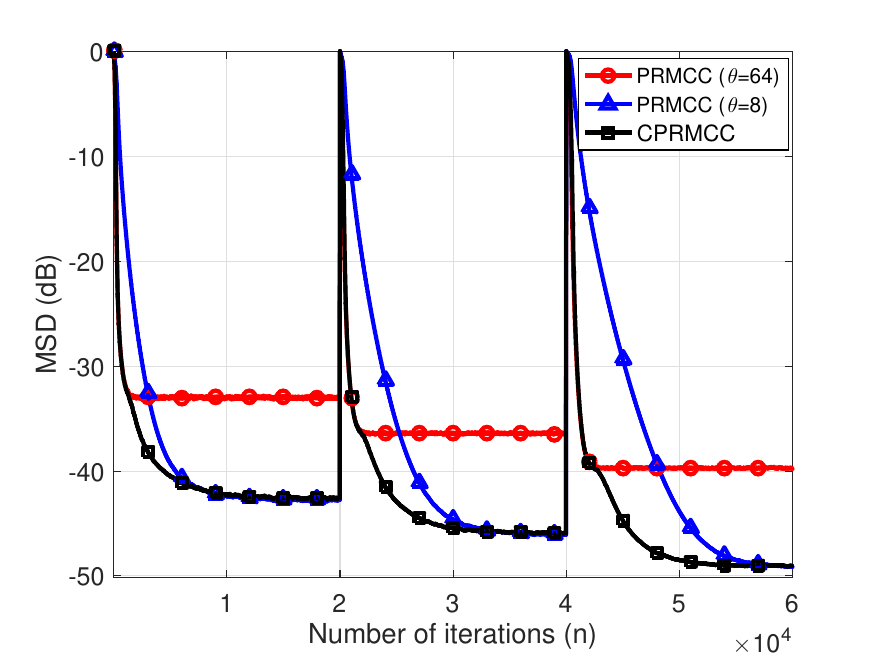}
   \caption{The MSD performance comparison between the CPRMCC and its component PRMCCs.}
   \label{Figure_summary2}
\end{figure}
%


In the second experiment, the advantage of CPRMCC over PRMCC was tested. The considered order-128 sparse system has 8, 16, and 32 non-zero taps at $n=0$, $n=20000$ and $n=40000$, respectively. The non–zero coefficients were generated as Gaussian variables with zero mean and unit variance with their positions randomly selected.
The observation noise has a distribution of $v(n)\sim0.99\mathcal{N}(0,10^{-4})+0.01\mathcal{N}(0,100)$.

The CPRMCC had the following parameter setting: $\lambda=0.99$, $\sigma=\sigma_b=2$, $\mu_b=50$, $b^+=4$, $\beta=0.999$ and $\gamma=2$. The trace controllers for its two component PRMCC filters were $\theta_1 = 64$ and $\theta_2=8$, respectively. Other parameters of the PRMCC filters were set the same as those in the first experiment.
The weight vector of the second component filter ${\bf w}_2(n)$ was updated according to \eqref{CPRMCC5}. In Fig.~\ref{Figure_summary2}, MSD comparison between the CPRMCC and component PRMCCs is shown. As expected, the PRMCC achieves a better steady-state MSD at the cost of slower convergence, and vice versa. The CPRMCC taking advantages from both component PRMCC filters, however, achieves fast convergence and low steady-state performance at the same time. It behaves at least as good as the better component PRMCC filter.

\begin{figure}[!ht] 
   \hspace{-0.4cm}%
   \centering
   \includegraphics[width=12cm,keepaspectratio]
   {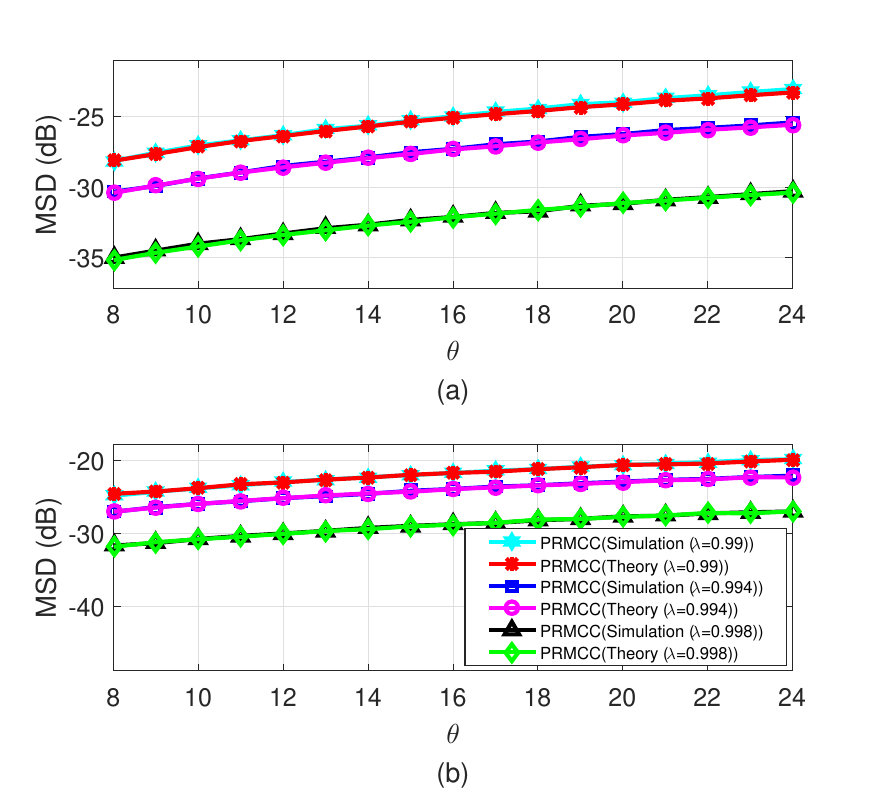}
   \caption{Steady-state MSDs of the proposed PRMCC algorithm: (a) mixed-Gaussian noise, (b) uniform noise.}
   \label{Figure_summary4}
\end{figure}

In the final experiment, we verify the accuracy of the derived theoretical steady-state MSD performance given in \eqref{SSA8} for PRMCC. Both stationary and nonstationary systems were considered and for each system, two types of observation noises were investigated: a mixed-Gaussian noise $v(n)\sim0.99\mathcal{N}(0,10^{-4})+0.01\mathcal{N}(0,400)$ and a uniform noise distributed over $[-0.5,0.5]$. The kernel bandwidth was set as $\sigma=1$.
First, the results for the identification of a stationary system ${\bf w}(0)=[0.7071,0,0,0,0,0,0,0.7071]$ are presented.
The empirical and analytical steady-state MSDs for different values of $\theta$ and $\lambda$ are compared in Fig.~\ref{Figure_summary4}. In all combinations of $\theta$ and $\lambda$, close match between simulated and derived MSDs is observed. Moreover, the steady-state MSD performance improves with a smaller $\theta$ and a larger $\lambda$, as expected.

\begin{figure}[!ht] 
   \hspace{-0.4cm}%
   \centering
   \includegraphics[width=12cm,keepaspectratio]
   {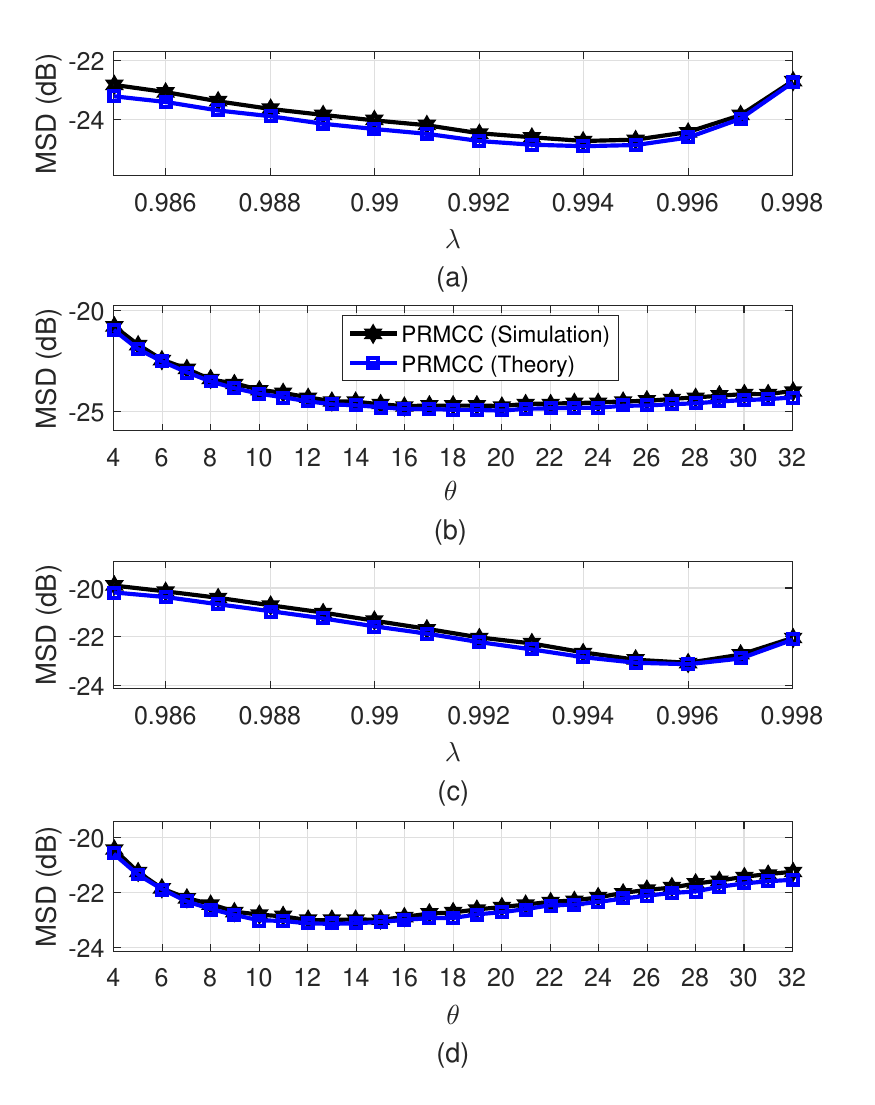}
   \caption{Tracking abilities of PRMCC evaluated against different $\lambda$ and $\theta$: (a) mixed-Gaussian noise ($\theta=16$), (b) mixed-Gaussian noise ($\lambda=0.995$), (c) uniform noise ($\theta=16$), (d) uniform noise ($\lambda=0.995$).}
   \label{Figure_summary5}
\end{figure}
\begin{table}[!ht]
\renewcommand{\arraystretch}{2}
\begin{center}
\caption{Optimal parameter values for a nonstationary system identification.}
\label{Table2}
{\begin{tabular}{|c||p{3.5cm}<{\centering}|p{2.3cm}<{\centering}|p{3.5cm}<{\centering}|p{3.5cm}<{\centering}|}\hline  {$v(n)$} & {mixed-Gaussian noise} &  {uniform noise}
\\\hline\hline {$\lambda_{opt} (\rm Simulation)$} & 0.994 & 0.996
\\\hline {$\lambda_{opt} (\rm Theory)$} & 0.993 & 0.9952
\\\hline {$\theta_{opt} (\rm Simulation)$} & 19 & 14
\\\hline {$\theta_{opt} (\rm Theory)$} & 21.238 & 15.392   \\\hline
\end{tabular}}{}
\end{center}
\end{table}

The nonstationary system followed the random walk model in \eqref{TA1}, for which $\sigma_q^2$ was chosen as $10^{-6}$.
The MSD comparison was shown in Fig.~\ref{Figure_summary5}, where slight mismatches between empirical and analytical curves are observed due to the introduction of noise ${\bf q}(n)$. From the figure, optimal combinations of $\theta$ and $\lambda$ leading to minimum steady-state MSDs were found and listed in Table~\ref{Table2}. The empirically optimal parameters generally match theoretically optimal parameters obtained via \eqref{TA6} and \eqref{TA7}.

\section{Conclusion}
\label{conclusion}

Motivated by recently proposed proportionate recursive least squares (PRLS) algorithm, a proportionate recursive maximum correntropy criterion (PRMCC) adaptive filtering algorithm was designed. To address the conflicting objectives of faster convergence and lower steady-state error, an adaptive convex combination of two PRMCC filters was investigated, leading to the combinational PRMCC  (CPRMCC) algorithm. Theoretical performance analyses were provided for the PRMCC and CPRMCC algorithms.
Numerical simulations of system identification verified the superiority of PRMCC/CPRMCC over the non-sparsity-aware RMCC, as well as the accuracy of theoretical performance results.


\begin{appendices}

\section{Proof of \eqref{Sta13}}
\label{Proof of first order performance of PRMCC}

In order to make the stability analysis tractable, we approximate ${\bf P}(n)$ and ${\bf G}(n-1)$ by their steady-state means ${\bf P}$ and ${\bf G}$.
We firstly derive the steady-state mean of ${\bf P}^{-1}(n)$.
According to \cite{Zhang16}, ${\bf P}^{-1}(n)$ can be written as
\begin{align}
    \label{Sta4}
    &{\bf P}^{-1}(n)\!=\!\lambda^{n+1}\delta {\bf I}_N+\sum_{i=0}^{n}\lambda^{n-i}{\rm{exp}}(-\frac{e^2(i|n)}{2\sigma^2}){\bf x}_N(i){\bf x}_N^T(i),
\end{align}
where $\delta$ is preset. As $\lambda<1$, the steady-state mean of ${\bf P}^{-1}(n)$ as $n$$\to$$\infty$, is given by
\begin{eqnarray}
    \label{Sta5}
    {\bf P}^{-1}=\lim\limits_{n\to\infty}\text{E}\left[{\bf P}^{-1}(n)\right]=\frac{\text{E}[{\rm{exp}}(-\frac{e^2(\infty|n)}{2\sigma^2})]{\bf R}}{1-\lambda},
\end{eqnarray}
in which ${\bf R}=\text{E}\left[{\bf x}_N(n){\bf x}_N^T(n)\right]$ is the input covariance matrix.

By applying $\lim\limits_{i\to\infty}e(i|n)=v(n)$ and the Taylor series expansion, $\text{E}[{\rm{exp}}(-\frac{e^2(\infty|n)}{2\sigma^2})]$ can be expressed as
\begin{eqnarray}
    \label{Sta6}
    \text{E}[{\rm{exp}}(-\frac{e^2(\infty|n)}{2\sigma^2})]=\text{E}[1-\frac{v^2(n)}{2\sigma^2}+\frac{v^4(n)}{8\sigma^4}+o(v^4(n))]\approx 1-\frac{\sigma_v^2}{2\sigma^2}+\frac{\text{E}[v^4(n)]}{8\sigma^4},
\end{eqnarray}
where the second approximation holds when the estimation noise power is sufficiently small such that the term $\text{E}[o(v^4(n))]$ is negligible.

Based on \eqref{Sta5} and \eqref{Sta6}, we can approximate the mean value of ${\bf P}(n)$ as
\begin{eqnarray}
    \label{Sta7}
    \text{E}\left[{\bf P}(n)\right]\approx \text{E}\left[{\bf P}^{-1}(n)\right]^{-1}={\bf P}\approx\frac{(1-\lambda)}{1-\frac{\sigma_v^2}{2\sigma^2}+\frac{\text{E}[v^4(n)]}{8\sigma^4}}{\bf R}^{-1}.
\end{eqnarray}

The $\text{E}\left[{\bf G}(n-1)\right]$ can be approximated by its steady-state mean ${\bf G}$, whose $i$-th diagonal element can be easily obtained from \eqref{PPRMCCAlgorithm10} as \cite{Qin20} 
\begin{eqnarray}
    \label{Sta8}
    g_i \approx \theta\left(\frac{1-\alpha}{2N}+(1+\alpha)\frac{|w_i|}{2||{\bf w}_N||_1}\right).
\end{eqnarray}

Regarding $f(e(n|n\!\!-\!\!1))$ in \eqref{PPRMCCAlgorithm61}, we have, again according to the Taylor series expansion, the following approximation
\begin{eqnarray}
    \label{Sta9}
    &&f(e(n|n\!\!-\!\!1))\nonumber\\
    &\!\!\!\approx\!\!\!& e(n|n\!\!-\!\!1)-\frac{e^3(n|n\!\!-\!\!1)}{2\sigma^2}+\frac{e^5(n|n\!\!-\!\!1)}{8\sigma^4}\nonumber\\
    &\!\!\!\approx\!\!\!& {\bf x}^T_N(n){\bf h}_N(n\!-\!1)+v(n)-\frac{v^3(n)+3{\bf x}^T_N(n){\bf h}_N(n\!-\!1)v^2(n)}{2\sigma^2}+ \frac{v^5(n)+5{\bf x}^T_N(n){\bf h}_N(n\!-\!1)v^4(n)}{8\sigma^4},
\end{eqnarray}
where
\begin{eqnarray}
    \label{Sta91}
    e(n|n\!-\!1)={\bf x}^T_N(n){\bf h}_N(n-1)+v(n),
\end{eqnarray}
and the approximation in the third line of \eqref{Sta9} is valid when the noise power is small. Substituting  \eqref{Sta9} into \eqref{Sta3} yields
\begin{eqnarray}
    \label{Sta10}
    {\bf h}_N(n)&\!\!\!=\!\!\!&[{\bf I}_N-{\bf A}(n)]{\bf h}_N(n\!-\!1)-{\bf a}(n),
\end{eqnarray}
where
\begin{eqnarray}
    \label{Sta11}
    \hspace{-0cm}{\bf A}(n)=(1-\frac{3v^2(n)}{2\sigma^2}+\frac{5v^4(n)}{8\sigma^4}){\bf G}(n\!-\!1){\bf P}(n){\bf x}_{N}(n){\bf x}^T_{N}(n),
\end{eqnarray}
\begin{eqnarray}
    \label{Sta111}
    \hspace{-0cm}{\bf a}(n)=(v(n)\!-\!\frac{v^3(n)}{2\sigma^2}\!+\!\frac{v^5(n)}{8\sigma^4}){\bf G}(n\!-\!1){\bf P}(n){\bf x}_{N}(n).
\end{eqnarray}

To obtain the stability condition for the PRMCC, we express \eqref{Sta10} as
\begin{eqnarray}
    \label{Sta12}
    {\bf h}_N(n)=\prod_{i=0}^{n-1}[{\bf I}_N-{\bf A}(n-i)]{\bf h}_N(0)-\prod_{i=0}^{n-2}[{\bf I}_N-{\bf A}(n-i)]{\bf a}(1)-\dots -{\bf a}(n),
\end{eqnarray}

Based on \eqref{Sta7}, \eqref{Sta8} and {\bf Assumptions 1-3},  we take the expectation on both sides of \eqref{Sta12} and obtain
\begin{align}
    \label{Sta13_another}
    &\text{E}[{\bf h}_N(n)]\approx\bigg[{\bf I}_N\!-\!\frac{(1\!-\!\lambda)(1\!-\!\frac{3\sigma_v^2}{2\sigma^2}\!+\!\frac{5\text{E}[v^4(n)]}{8\sigma^4})}
    {1-\frac{\sigma_v^2}{2\sigma^2}+\frac{\text{E}[v^4(n)]}{8\sigma^4}}{\bf G}\bigg]^n\text{E}[{\bf h}_N(0)].
\end{align}

\section{Proof of \eqref{SSA7}}
\label{proof of second order mean square performance of PRMCC}

For convenience, we rewrite \eqref{SSA1} as
\begin{eqnarray}
    \label{SSA1_another}
    {\bf K}(n)&\!\!\!=\!\!\!&{\bf K}(n-1)-\text{E}[{\bf h}_N(n\!-\!1){\bf x}^T_{N}(n){\bf P}(n){\bf G}(n\!-\!1) f(e(n|n\!-\!1))]-\text{E}[{\bf G}(n\!-\!1){\bf P}(n){\bf x}_{N}(n){\bf h}^T_N(n\!-\!1)f(e(n|n\!-\!1))]\nonumber\\
    &&+\text{E}[{\bf G}(n\!-\!1){\bf P}(n){\bf x}_{N}(n){\bf x}^T_{N}(n){\bf P}(n){\bf G}(n\!-\!1)f^2(e(n|n\!-\!1))],
\end{eqnarray}
By using \eqref{Sta9} and {\bf Assumptions 1-2}, the second term on the RHS in \eqref{SSA1_another} can be simplified as
\begin{align}
    \label{SSA2}
&\text{E}[{\bf h}_N(n\!-\!1){\bf x}^T_{N}(n){\bf P}(n){\bf G}(n\!-\!1)f(e(n|n\!-\!1))]\nonumber\\
&\approx\text{E}[{\bf h}_N(n\!-\!1){\bf h}^T_N(n\!-\!1){\bf x}_{N}(n){\bf x}^T_{N}(n){\bf P}(n){\bf G}(n\!-\!1)\nonumber\\
    &-\frac{3v^2(n){\bf h}_N(n\!-\!1){\bf h}^T_N(n\!-\!1){\bf x}_{N}(n){\bf x}^T_{N}(n){\bf P}(n){\bf G}(n\!-\!1)}{2\sigma^2}\nonumber\\
    &+\frac{5v^4(n){\bf h}_N(n\!-\!1){\bf h}^T_N(n\!-\!1){\bf x}_{N}(n){\bf x}^T_{N}(n){\bf P}(n){\bf G}(n\!-\!1)}{8\sigma^4}]\nonumber\\
    &\approx\frac{(1\!-\!\lambda)(1\!-\!\frac{3\sigma_v^2}{2\sigma^2}\!+\!\frac{5\text{E}[v^4(n)]}{8\sigma^4})}
    {1-\frac{\sigma_v^2}{2\sigma^2}+\frac{\text{E}[v^4(n)]}{8\sigma^4}}{\bf K}(n-1){\bf G}.
\end{align}
In a similar way, we obtain the approximation of the third term in the following, which is given by
\begin{align}
    \label{SSA3}
    \text{E}[{\bf G}(n\!-\!1){\bf P}(n){\bf x}_{N}(n){\bf h}^T_N(n\!-\!1)f(e(n|n\!-\!1))]\approx\frac{(1\!-\!\lambda)(1\!-\!\frac{3\sigma_v^2}{2\sigma^2}\!+\!\frac{5\text{E}[v^4(n)]}{8\sigma^4})}
    {1-\frac{\sigma_v^2}{2\sigma^2}+\frac{\text{E}[v^4(n)]}{8\sigma^4}}{\bf G}{\bf K}(n-1).
\end{align}

We continue to use the Taylor series expansion and \eqref{Sta91} to expand $f^2(e(n|n\!-\!1))$ as
\begin{eqnarray}
    \label{SSA4}
    f^2(e(n|n\!-\!1))&=&{\rm{exp}}(-\frac{e^2(n|n\!-\!1)}{\sigma^2})e^2(n|n\!-\!1)\nonumber\\
    &\approx& e^2(n|n\!-\!1)-\frac{e^4(n|n\!-\!1)}{\sigma^2}+\frac{e^6(n|n\!-\!1)}{2\sigma^4}\nonumber\\
    &\approx& v^2(n)+2{\bf x}^T_N(n){\bf h}_N(n\!-\!1)v(n)+{\bf x}^T_N(n){\bf h}_N(n\!-\!1){\bf h}^T_N(n\!-\!1){\bf x}_N(n)\nonumber\\
    &&-\frac{v^4(n)+4v^3(n){\bf x}^T_N(n){\bf h}_N(n\!-\!1)+6v^2(n){\bf x}^T_N(n){\bf h}_N(n\!-\!1){\bf h}^T_N(n\!-\!1){\bf x}_N(n)}{\sigma^2}\nonumber\\
    &&+\frac{v^6(n)+6v^5(n){\bf x}^T_N(n){\bf h}_N(n\!-\!1)+15v^4(n){\bf x}^T_N(n){\bf h}_N(n\!-\!1){\bf h}^T_N(n\!-\!1){\bf x}_N(n)}{2\sigma^4},
\end{eqnarray}
such that the fourth term in \eqref{SSA1_another} can be evaluated as 
%
%
\begin{align}
    \label{SSA5}
    &\text{E}[{\bf G}(n\!-\!1){\bf P}(n){\bf x}_{N}(n){\bf x}^T_{N}(n){\bf P}(n){\bf G}(n\!-\!1)f^2(e(n|n\!-\!1))]\nonumber\\
    &\approx \frac{(1-\lambda)^2}{\sigma_x^4(1-\frac{\sigma_v^2}{2\sigma^2}+\frac{\text{E}[v^4(n)]}{8\sigma^4})^2}{\bf G}\cdot\text{E}\bigg[(v^2(n)-\frac{v^4(n)}{\sigma^2}+\frac{v^6(n)}{2\sigma^4}){\bf x}_N(n){\bf x}^T_N(n)\nonumber\\
    &+(1-\frac{6v^2(n)}{\sigma^2}+\frac{15v^4(n)}{2\sigma^4}){\bf x}_N(n){\bf x}^T_N(n){\bf h}_N(n\!-\!1){\bf h}^T_N(n\!-\!1){\bf x}_N(n){\bf x}^T_N(n)\bigg]{\bf G}\nonumber\\
    &\approx \frac{(1-\lambda)^2(\sigma_v^2-\frac{\text{E}[v^4(n)]}{\sigma^2}+\frac{\text{E}[v^6(n)]}{2\sigma^4})}
    {\sigma_x^2(1-\frac{\sigma_v^2}{2\sigma^2}+\frac{\text{E}[v^4(n)]}{8\sigma^4})^2}{\bf G}^2+\frac{(1-\lambda)^2(1-\frac{6\sigma_v^2}{\sigma^2}+\frac{15\text{E}[v^4(n)]}{2\sigma^4})}
    {(1-\frac{\sigma_v^2}{2\sigma^2}+\frac{\text{E}[v^4(n)]}{8\sigma^4})^2}{\bf G}{\bf K}(n-1){\bf G},
\end{align}
where the first approximation is due to {\bf Assumption 3}, and the second approximation is obtained by replacing ${\bf x}_N(n){\bf x}_N^T(n)$ with ${\bf R}$ and noting that ${\bf R}^{-1}=\frac{1}{\sigma_x^2}{\bf I}_N$. 
Substituting \eqref{SSA2}, \eqref{SSA3} and \eqref{SSA5} into \eqref{SSA1_another} and letting $n$$\to$$\infty$, we further arrive at
\begin{align}
    \label{SSA6}
    &(1\!-\!\frac{3\sigma_v^2}{2\sigma^2}\!+\!\frac{5\text{E}[v^4(n)]}{8\sigma^4})({\bf K}(\infty){\bf G}+{\bf G}{\bf K}(\infty))-\frac{(1-\lambda)(1-\frac{6\sigma_v^2}{\sigma^2}+\frac{15\text{E}[v^4(n)]}{2\sigma^4})}
    {1-\frac{\sigma_v^2}{2\sigma^2}+\frac{\text{E}[v^4(n)]}{8\sigma^4}}{\bf G}{\bf K}(\infty){\bf G}\nonumber\\
    &\approx \frac{(1-\lambda)(\sigma_v^2-\frac{\text{E}[v^4(n)]}{\sigma^2}+\frac{\text{E}[v^6(n)]}{2\sigma^4})}
    {\sigma_x^2(1-\frac{\sigma_v^2}{2\sigma^2}+\frac{\text{E}[v^4(n)]}{8\sigma^4})}{\bf G}^2,
\end{align}
from which the $(i,i)$-th element  of ${\bf K}(\infty)$ is equal to 
\begin{align}
    \label{SSA7_another}
    {\bf K}_i(\infty) \approx \frac{\frac{(1-\lambda)g_i(\sigma_v^2-\frac{\text{E}[v^4(n)]}{\sigma^2}+\frac{\text{E}[v^6(n)]}{2\sigma^4})}
    {\sigma_x^2(1-\frac{\sigma_v^2}{2\sigma^2}+\frac{\text{E}[v^4(n)]}{8\sigma^4})}}{2\!-\!\frac{3\sigma_v^2}{\sigma^2}
    \!+\!\frac{5\text{E}[v^4(n)]}{4\sigma^4}-\frac{(1-\lambda)g_i(1-\frac{6\sigma_v^2}{\sigma^2}+\frac{15\text{E}[v^4(n)]}{2\sigma^4})}
    {1-\frac{\sigma_v^2}{2\sigma^2}+\frac{\text{E}[v^4(n)]}{8\sigma^4}}}.
\end{align}

\section{Proof of \eqref{CPRMCC18}}
\label{Proof of combination of two PRMCC in the steady state}

Based on \eqref{Sta3}, the weight error vectors of both component filters are given by
\begin{eqnarray}
    \label{CPRMCC10}
    {\bf h}_1(n)\!=\!{\bf h}_1(n\!\!-\!\!1)\!-\!{\bf G}_1(n\!\!-\!\!1){\bf P}_1(n){\bf x}_{N}(n)f(e_1(n|n\!\!-\!\!1)),\\
    \label{CPRMCC11}
    {\bf h}_2(n)\!=\!{\bf h}_2(n\!\!-\!\!1)\!-\!{\bf G}_2(n\!\!-\!\!1){\bf P}_2(n){\bf x}_{N}(n)f(e_2(n|n\!\!-\!\!1)).
\end{eqnarray}
Taking the outer product of ${\bf h}_1(n)$ and ${\bf h}_2(n)$ then applying the statistical expectation, we can obtain
\begin{eqnarray}
    \label{CPRMCC12}
    {\bf H}_{12}(n)&\!\!\!=\!\!\!&{\bf H}_{12}(n-1)-\text{E}[{\bf h}_1(n-1){\bf x}^T_{N}(n){\bf P}_2(n) {\bf G}_2(n-1)f(e_2(n|n-1))]\nonumber\\
    &&-\text{E}[{\bf G}_1(n-1){\bf P}_1(n){\bf x}_{N}(n){\bf h}^T_2(n-1)f(e_1(n|n-1))]\nonumber\\
    &&+\text{E}[{\bf G}_1(n-1){\bf P}_1(n){\bf x}_{N}(n){\bf x}^T_{N}(n){\bf P}_2(n){\bf G}_2(n\!-\!1)f(e_1(n|n\!-\!1))f(e_2(n|n\!-\!1))].
\end{eqnarray}

Similar to the derivations of \eqref{SSA2} and \eqref{SSA3}, the second and third terms on the RHS of \eqref{CPRMCC12} can be computed as
\begin{align}
    \label{CPRMCC13}
    \text{E}[{\bf h}_1(n\!-\!1){\bf x}^T_{N}(n){\bf P}_2(n){\bf G}_2(n\!-\!1)f(e_2(n|n\!-\!1))]\approx\frac{(1\!-\!\lambda)(1\!-\!\frac{3\sigma_v^2}{2\sigma^2}\!+\!\frac{5\text{E}[v^4(n)]}{8\sigma^4})}
    {1-\frac{\sigma_v^2}{2\sigma^2}+\frac{\text{E}[v^4(n)]}{8\sigma^4}}{\bf H}_{12}(n-1){\bf G}_2,
\end{align}
\begin{align}
    \label{CPRMCC14}
    \text{E}[{\bf G}_1(n\!-\!1){\bf P}_1(n){\bf x}_{N}(n){\bf h}^T_2(n\!-\!1)f(e_1(n|n\!-\!1))]\approx\frac{(1\!-\!\lambda)(1\!-\!\frac{3\sigma_v^2}{2\sigma^2}\!+\!\frac{5\text{E}[v^4(n)]}{8\sigma^4})}
    {1-\frac{\sigma_v^2}{2\sigma^2}+\frac{\text{E}[v^4(n)]}{8\sigma^4}}{\bf G}_1{\bf H}_{12}(n-1),
\end{align}
where the $i$-th diagonal element of ${\bf G}_j$, $j=1,2$, is
\begin{eqnarray}
    \label{CPRMCC15}
    g_{j,i} \approx \theta_j t_i= \theta_j\left(\frac{1-\alpha}{2N}+(1+\alpha)\frac{|w_i|}{2||{\bf w}_N||_1}\right).
\end{eqnarray}

Before analyzing the fourth term of RHS in \eqref{CPRMCC12}, we firstly derive the approximation of $f(e_1(n|n\!-\!1))f(e_2(n|n\!-\!1))$
\begin{eqnarray}
    \label{CPRMCC16}
    &&f(e_1(n|n\!-\!1)f(e_2(n|n\!-\!1)\nonumber\\
    &\approx& (e_1(n|n-1)-\frac{e_1^3(n|n-1)}{2\sigma^2})(e_2(n|n-1)-\frac{e_2^3(n|n-1)}{2\sigma^2})\nonumber\\
    &=&e_1(n|n-1)e_2(n|n-1)-\frac{e_1(n|n-1)e_2^3(n|n-1)+e_1^3(n|n-1)e_2(n|n-1)}{2\sigma^2}+\frac{e_1^3(n|n-1)e_2^3(n|n-1)}{4\sigma^4}\nonumber\\
    &\approx& {\bf x}_{N}^T(n){\bf h}_1(n-1){\bf h}_2^T(n-1){\bf x}_{N}(n)+{\bf x}_{N}^T(n)({\bf h}_1(n-1)+{\bf h}_2(n-1))v(n)+v^2(n)\nonumber\\
    &&-\frac{2v^4(n)+4{\bf x}_{N}^T(n)({\bf h}_1(n-1)+{\bf h}_2(n-1))v^3(n)+6{\bf x}_{N}^T(n){\bf h}_1(n-1){\bf h}_2^T(n-1){\bf x}_{N}(n)v^2(n)}{2\sigma^2}\nonumber\\
    &&+\frac{v^6(n)+3{\bf x}_{N}^T(n)({\bf h}_1(n-1)+{\bf h}_2(n-1))v^5(n)+9{\bf x}_{N}^T(n){\bf h}_1(n-1){\bf h}_2^T(n-1){\bf x}_{N}(n)v^4(n)}{4\sigma^4},
\end{eqnarray}
$\rm where$ $e_i(n|n-1)={\bf x}_{N}^T(n){\bf h}_i(n-1)+v(n)$ $\rm and$ $e_i^3(n|n-1)\approx 3{\bf x}_{N}^T(n){\bf h}_i(n-1)v^2(n)+v^3(n)$, $i=1,2$.

Now we have
\begin{eqnarray}
    \label{CPRMCC17}
    &&\text{E}[{\bf G}_1(n-1){\bf P}_1(n){\bf x}_{N}(n){\bf x}^T_{N}(n){\bf P}_2(n){\bf G}_2(n\!-\!1)f(e_1(n|n\!-\!1))f(e_2(n|n\!-\!1))]\nonumber\\
    &\approx& \frac{(1-\lambda)^2}{\sigma_x^4(1-\frac{\sigma_v^2}{2\sigma^2}+\frac{\text{E}[v^4(n)]}{8\sigma^4})^2}{\bf G}_1\text{E}\bigg[(v^2(n)-\frac{v^4(n)}{\sigma^2}+\frac{v^6(n)}{4\sigma^4}){\bf x}_N(n){\bf x}^T_N(n)\nonumber\\
    &&+(1-\frac{3v^2(n)}{\sigma^2}+\frac{9v^4(n)}{4\sigma^4}){\bf x}_N(n){\bf x}^T_N(n){\bf h}_1(n\!-\!1){\bf h}^T_2(n\!-\!1){\bf x}_N(n){\bf x}^T_N(n)\bigg]{\bf G}_2\nonumber\\
    &\approx& \frac{(1-\lambda)^2(\sigma_v^2-\frac{\text{E}[v^4(n)]}{\sigma^2}+\frac{\text{E}[v^6(n)]}{4\sigma^4})}
    {\sigma_x^2(1-\frac{\sigma_v^2}{2\sigma^2}+\frac{\text{E}[v^4(n)]}{8\sigma^4})^2}{\bf G}_1{\bf G}_2+\frac{(1-\lambda)^2(1-\frac{3\sigma_v^2}{\sigma^2}+\frac{9\text{E}[v^4(n)]}{4\sigma^4})}
    {(1-\frac{\sigma_v^2}{2\sigma^2}+\frac{\text{E}[v^4(n)]}{8\sigma^4})^2}{\bf G}_1{\bf H}_{12}(n-1){\bf G}_2.
\end{eqnarray}

With \eqref{CPRMCC82}, \eqref{CPRMCC13}, \eqref{CPRMCC14} and \eqref{CPRMCC17}, we finally arrive at
\begin{align}
    \label{CPRMCC18_another}
    {\rm MSD}_{12}\approx \sum_{i=1}^{N}\frac{\frac{(1-\lambda)\theta_{12}t_i(\sigma_v^2-\frac{\text{E}[v^4(n)]}{\sigma^2}+\frac{\text{E}[v^6(n)]}{4\sigma^4})}
    {\sigma_x^2(1-\frac{\sigma_v^2}{2\sigma^2}+\frac{\text{E}[v^4(n)]}{8\sigma^4})}}{(1\!-\!\frac{3\sigma_v^2}{2\sigma^2}
    \!+\!\frac{5\text{E}[v^4(n)]}{8\sigma^4})-\frac{(1-\lambda)\theta_{12}t_i(1-\frac{3\sigma_v^2}{\sigma^2}+\frac{9\text{E}[v^4(n)]}{4\sigma^4})}
    {1-\frac{\sigma_v^2}{2\sigma^2}+\frac{\text{E}[v^4(n)]}{8\sigma^4}}},
\end{align}
with
\begin{eqnarray}
    \label{CPRMCC19_1}
    \theta_{12}=\frac{\theta_1\theta_2}{\theta_1+\theta_2}.
\end{eqnarray}

\end{appendices}

\section*{Acknowledgements}
This work was supported in part by the National Natural Science Foundation of China under Grants No. 62271138, No. 61871114, and No. U20B2039; in part by the Fund for returning students to study abroad in Nanjing; in
part by the Priority Academic Program Development (PAPD) of Jiangsu Higher Education Institutions, and in part by the Fundamental Research Funds for the Central Universities under Grant No. 2242021k30019.



\bibliographystyle{elsarticle-num}

\end{document}